\documentclass[superscriptaddress,secnumarabic,
amssymb,amsmath,nobibnotes,aps,prd,showkeys,nofootinbib,twocolumn]{revtex4}%
\usepackage{graphicx}
\usepackage{epsf}
\usepackage{bm}
\usepackage{amsmath}
\usepackage{amsfonts}
\usepackage{amssymb}
\usepackage{epstopdf}
\usepackage{natbib}%
\usepackage{color}

\begin{document}

\title{Jeans instability criterion from the viewpoint of non-gaussian statistics}
\author{Everton M. C. Abreu}\email{evertonabreu@ufrrj.br}
\affiliation{Grupo de F\' isica Te\'orica e Matem\'atica F\' isica, Departamento de F\'{i}sica, Universidade Federal Rural do Rio de Janeiro, 23890-971, Serop\'edica, RJ, Brazil}
\affiliation{Departamento de F\'{i}sica, Universidade Federal de Juiz de Fora, 36036-330, Juiz de Fora, MG, Brazil}
\author{Jorge Ananias Neto}\email{jorge@fisica.ufjf.br}
\affiliation{Departamento de F\'{i}sica, Universidade Federal de Juiz de Fora, 36036-330, Juiz de Fora, MG, Brazil}
\author{Edesio M. Barboza Jr.} \email{edesiobarboza@uern.br}
\affiliation{Departamento de F\'isica, Universidade do Estado do Rio Grande do Norte, 59610-210, Mossor\'o, RN, Brazil}
\author{Rafael C. Nunes}\email{nunes@ecm.ub.edu}
\affiliation{Funda\c{c}\~ao CAPES, Minist\'erio da Educa\c{c}\~ao e Cultura, 70040-020, Bras\' ilia, DF, Brazil}

\begin{abstract}
\noindent In this Letter we have derived the Jeans length in the context of the Kaniadakis statistics. 
We have compared this result with the Jeans length obtained in the non-extensive Tsallis statistics 
and discussed the main differences between these two models. We have also obtained the $\kappa$-sound velocity.  
Finally, we have applied the results obtained here to analyze an astrophysical system.

\end{abstract}

\maketitle

The dynamical stability of a self-gravitating system usually can be described by the Jeans criterion of gravitational instability. The so-called Jeans's length \cite{j1} is given by

\begin{eqnarray}
\label{lc}
\lambda_J=\sqrt{\frac{\pi k_B T}{\mu m_H G \rho_0}},
\end{eqnarray}
where $k_B$ is the Boltzmann constant, $T$ is the temperature, $\mu$ is the mean molecular weight, $m_H$ is the atomic mass of hydrogen, $G$ is the gravitational constant and $\rho_0$ is the equilibrium mass density. The critical value $\lambda_J$, Eq.(\ref{lc}), is derived by considering a small perturbation in a set of four equations that are the equation of continuity, the Euler's equation, the Poisson's equation and the equation of state of an ideal gas. The Jeans length establishes that if the wave length $\lambda$ of density fluctuation is greater than $\lambda_J$ then 
the density will grow with time in an exponential form and the system will become gravitationally unstable.
For more details see Jiulin in ref. \cite{DJ}.

Tsallis \cite{tsallis} has proposed an important extension of the Boltzman-Gibbs (BG) statistical theory.   In a brief and technical terminology, this model is also currently referred to as a nonextensive (NE) statistical mechanics. 

Tsallis thermostatistics formalism defines a nonadditive entropy as

\begin{eqnarray}
\label{nes}
S_q =  k_B \, \frac{1 - \sum_{i=1}^W p_i^q}{q-1}\;\;\;\;\;\;\qquad \Big(\,\sum_{i=1}^W p_i = 1\,\Big)\,\,,
\end{eqnarray}
where $p_i$ is the probability of the system to be in a microstate, $W$ is the total number of configurations and 
$q$, known in the current literature as being the Tsallis parameter or NE  parameter, is a real parameter which quantifies the degree of nonextensivity. 
The definition of entropy in Tsallis statistics carries the usual properties of positivity, equiprobability, concavity and irreversibility and it also has motivated the study of multifractals systems.
It is important to stress that Tsallis thermostatistics formalism contains the BG statistics as a particular case in the limit $ q \rightarrow 1$ where the usual additivity of entropy is recovered. 

Plastino and Lima \cite{PL} have derived a NE equipartition law of energy. It has been shown that the kinetic foundations of Tsallis' NE statistics leads to a velocity distribution for free particles given by  \cite{SPL}

\begin{eqnarray}
\label{vd}
f_q(v)=B_q \Big[1-(1-q) \frac{m v^2}{2 k_B T}\Big]^{1/1-q},
\end{eqnarray}
where $B_q$ is a normalization constant. Then, the expectation value of $v^2$, for each degree of freedom, is given by \cite{SA}

\begin{eqnarray}
\label{v2}
<v^2>_q=\frac{\int^\infty_0\, f_q \,v^2 dv}{\int^\infty_0\, f_q dv}\\ \nonumber
=\frac{2}{5-3q}\, \frac{k_B T}{m}.
\end{eqnarray}

The equipartition theorem is then obtained by using that

\begin{eqnarray}
\label{reqq}
E_q=\frac{1}{2} N m <v^2>_q,
\end{eqnarray}
and we will arrive at

\begin{eqnarray}
\label{ge}
E_q = \frac{1}{5 - 3 q} N k_B T\,\,.
\end{eqnarray}

The range of $q$ is $ 0 \le q < 5/3 $.  For $ q=5/3$ (critical value) the expression of the equipartition law of energy, Eq. (\ref{ge}), diverges. 
It is also easy to observe that for $ q = 1$,  the classical equipartition theorem for each microscopic degrees of freedom can be recovered.


On the other hand, concerning entropy, the recent formalism proposed by E. Verlinde  \cite{verlinde} obtains the gravitational acceleration  by using the holographic principle and the well known equipartition law of energy. His ideas relied on the fact that gravitation can be considered universal and independent of the details of the spacetime microstructure.  Besides, he brought new concepts concerning holography since the holographic principle must unify matter, gravity and quantum mechanics.

The model considers a spherical surface as being the holographic screen, with a particle of mass $M$ positioned in its center. The holographic screen can be imagined as a storage device for information. The number of bits, which is the smallest unit of information in the holographic screen, is assumed to be proportional to the  holographic screen
area $A$ and can be written
\begin{eqnarray}
\label{bits}
N = \frac{A }{\ell_P^2},
\end{eqnarray}
where $ A = 4 \pi r^2 $ and $\ell_P = \sqrt{G\hbar / c^3}$ is the Planck length and $\ell_P^2$ is the Planck area.   In Verlinde's framework one can suppose that the bits total energy on the screen is given by the equipartition law of energy

\begin{eqnarray}
\label{eq}
E = \frac{1}{2}\,N k_B T.
\end{eqnarray}

It is important to notice that the usual equipartition theorem in Eq. (\ref{eq}), can be derived from the usual BG thermostatistics.  Let us consider that the energy of the particle inside the holographic screen is equally divided by all bits in such a manner that we can have the expression

\begin{eqnarray}
\label{meq}
M c^2 = \frac{1}{2}\,N k_B T.
\end{eqnarray}

To obtain the gravitational acceleration, we can use  Eq. (\ref{bits}) and the Unruh temperature equation  \cite{unruh} given by

\begin{eqnarray}
\label{un}
k_B T = \frac{1}{2\pi}\, \frac{\hbar a}{c}\,\,.
\end{eqnarray}

Hence, we are  able to obtain the  (absolute) gravitational acceleration formula

\begin{eqnarray}
\label{acc}
a &=&  \frac{l_P^2 c^3}{\hbar} \, \frac{ M}{r^2}\nonumber\\ 
&=& G \, \frac{ M}{r^2}\,\,.
\end{eqnarray}
From Eq. (\ref{acc}) we can see that the Newton constant $G$ can be written in terms of the fundamental constants, $G=\ell_P^2 c^3/\hbar$.


As an application of NE equipartition theorem in Verlinde's formalism we can
use the NE equipartition formula, i.e., Eq. (\ref{ge}).  Hence, we can obtain a modified acceleration formula given by  \cite{abreu}
\begin{eqnarray}
\label{accm}
a = G_q \, \frac{ M}{r^2},
\end{eqnarray}
where $G_q$ is an effective gravitational constant which is written as

\begin{eqnarray}
\label{S}
G_q=\,\frac{5-3q}{2}\,G\,\,.
\end{eqnarray}

From result (\ref{S}) we can observe that the effective gravitational constant depends on the NE parameter $q$. For example, when $q=1$ we have $ G_q=G$ (BG scenario) and for $q\,=\,5 / 3$ we have the curious and hypothetical result 
$G_q=0$.  This result shows us that $q\,=\,5/3$ is an upper bound limit when we are dealing with the holographic screen, as we have said before.  But now we have other reasons to justify this classification.

Substituting $G$ by $G_q$, Eq.(\ref{S}), in (\ref{lc}), we can derive the same expression obtained by Jiulin\cite{DJ} for the NE critical wavelength

\begin{eqnarray}
\label{lbdaq}
\lambda_c^q=\sqrt{\frac{2}{5-3q} \frac{\pi k_B T}{\mu m_H G \rho_0}}=\sqrt{\frac{2}{5-3q}} \; \lambda_J.
\end{eqnarray}

This NE modification of the Jeans criterion leads to a new critical length $\lambda_c^q$ that depends on the nonextensive $q$-parameter as follows:

\vskip .2cm
\noindent - If $q=1 \Rightarrow \lambda>\lambda_c^q=\lambda_J$, the usual Jeans criterion is recovered.
\vskip .1 cm
\noindent - If $0<q<1$ the Jeans criterion is modified as $\lambda>\lambda_c^q<\lambda_J$.
\vskip .1 cm
\noindent - If $1<q<5/3$,  the Jeans criterion is modified as $\lambda>\lambda_c^q>\lambda_J$.
\vskip .1 cm
\noindent - If $q\rightarrow 5/3, \lambda_c^q \rightarrow \infty$, the self-gravitating system is always stable. 

Therefore, we have used Verlinde's formalism and Tsallis' thermostatistics in order to derive the NE Jeans length.

\vskip .2 cm

On the other hand, considering the well known Kaniadakis statistics \cite{kani1}, also called $\kappa$-statistics, similarly to Tsallis thermostatistics formalism generalizes the standard BG statistics initially by the introduction of $\kappa$-exponential and $\kappa$-logarithm defined by

\begin{eqnarray}
\label{expk}
exp_\kappa(f)=\Big( \sqrt{1+\kappa^2 f^2}+\kappa f \Big)^\frac{1}{\kappa},
\end{eqnarray}

\begin{eqnarray}
\label{logk}
\ln_\kappa(f)=\frac{f^\kappa-f^{-\kappa}}{2\kappa},
\end{eqnarray}
with the following operation being satisfied
\begin{eqnarray}
\ln_\kappa\Big(exp_\kappa(f)\Big)=exp_\kappa\Big(\ln_\kappa(f)\Big)\equiv f.
\end{eqnarray}
By Eqs. (\ref{expk}) and (\ref{logk}) we can observe that the $\kappa$-parameter deforms the usual definitions of the exponential and logarithm functions.

The $\kappa$-entropy associated with this $\kappa$-framework is given by
\begin{eqnarray}
S_\kappa=- k_B \sum_i^W  \,\frac{p_i^{1+\kappa}-p_i^{1-\kappa}}{2\kappa},
\end{eqnarray}
which recovers the BG entropy in the limit $\kappa \rightarrow 0$. It is important to mention here that the $\kappa$-entropy has satisfied the properties of concavity, additivity and extensivity. Tsallis' entropy satisfies the property of concavity and extensivity but not additivity. This property is not fundamental, in principle. The $\kappa$-statistics has been successfully applied in many experimental fronts. As an example we can mention cosmic rays \cite{Kanisca1} and cosmic effects \cite{aabn-1}, quark-gluon plasma \cite{Tewe}, kinetic models describing a gas of interacting atoms and photons \cite{Ross} and financial models \cite{RBJ}.

The kinetic foundations for the $\kappa$-statistics lead to a velocity distribution for free particles given by \cite{BSS}

\begin{eqnarray}
f_\kappa(v)=\Big[ \sqrt{1+\kappa^2 \Big( -\frac{m v^2}{2 k_B T}\Big)^2}  - \frac{\kappa m v^2}{2 k_B T} \Big]^\frac{1}{\kappa}.
\end{eqnarray}
Then, the expectation value of $v^2$, for each degree of freedom, is given by
\begin{eqnarray}
\label{vk}
<v^2>_\kappa=\frac{\int^\infty_0\, f_\kappa \,v^2 dv}{\int^\infty_0\, f_\kappa dv}.
\end{eqnarray}

Using the integral relation \cite{kani1}

\begin{eqnarray}
\int_0^\infty x^{r-1}\, exp_\kappa (-x) dx
=\,\frac{|2\kappa|^{-r}}{1+r|\kappa|}\; \frac{\Gamma\Big(\frac{1}{\left|2\kappa\right|}-\frac{r}{2}\Big)}{\Gamma\Big
(\frac{1}{\left|2\kappa\right|}+\frac{r}{2}\Big)}\; \Gamma(r)\,\,, \nonumber \\
\end{eqnarray}

we have that

\begin{eqnarray}
\label{vk2}
<v^2>_\kappa=\frac{(1+\frac{\kappa}{2})}{(1+\frac{3}{2}\kappa) \,2\kappa}
\frac{\Gamma{(\frac{1}{2\kappa}-\frac{3}{4})\,\Gamma{(\frac{1}{2\kappa}+\frac{1}{4})}}}{\Gamma{(\frac{1}{2\kappa}+\frac{3}{4})\,\Gamma{(\frac{1}{2\kappa}-\frac{1}{4})}}}\;  \frac{k_B T}{m}\,\,.  \nonumber \\
\end{eqnarray}

The $\kappa$-equipartition theorem can be obtained by using that

\begin{eqnarray}
\label{reqk}
E_\kappa=\frac{1}{2} N m <v^2>_\kappa\,,
\end{eqnarray}
where we arrive at\cite{gravo}

\begin{eqnarray}
\label{keq}
E_\kappa=\frac{1}{2} N \;\; \frac{(1+\frac{\kappa}{2})}{(1+\frac{3}{2}\kappa) \,2\kappa}
\frac{\Gamma{(\frac{1}{2\kappa}-\frac{3}{4})\,\Gamma{(\frac{1}{2\kappa}+\frac{1}{4})}}}{\Gamma{(\frac{1}{2\kappa}+\frac{3}{4})\,\Gamma{(\frac{1}{2\kappa}-\frac{1}{4})}}}\;\;\; k_B T\,\,. \nonumber \\
\end{eqnarray}

The range of $\kappa$ is $ 0 \le \kappa < 2/3 $.  For $ \kappa=2/3$ (critical value) the expression of the equipartition law of energy, Eq. (\ref{keq}), diverges, which is equivalent to $q=5/3$ in Tsallis formalism. For $ \kappa = 0$, the classical equipartition theorem for each microscopic degrees of freedom can be recovered. 


On the other hand, if we use the Kaniadakis equipartition theorem, Eq.(\ref{keq}), in the Verlinde formalism, the modified acceleration formula is given by
\begin{eqnarray}
\label{acck}
a = G_{\kappa} \, \frac{ M}{r^2}\,\,,
\end{eqnarray}
where $G_{\kappa}$ is an effective gravitational constant which is written as

\begin{eqnarray}
\label{Gk}
G_{\kappa}=  \frac{(1+\frac{3}{2}\kappa) \,2\kappa}{(1+\frac{\kappa}{2})}\; \;
\frac{\Gamma{(\frac{1}{2\kappa}+\frac{3}{4})\,\Gamma{(\frac{1}{2\kappa}-\frac{1}{4})}}}{\Gamma{(\frac{1}{2\kappa}-\frac{3}{4})\,\Gamma{(\frac{1}{2\kappa}+\frac{1}{4})}} }\; \; G
\end{eqnarray}
which can be written alternatively as $G=f(\kappa)\,G_{\kappa}$ used in Eq. (4.10)-(4.13), as we have explained in Section IV.

From the limits $\kappa=0, q=1$, with which we can recover the BG statistics, and $\kappa=2/3, q=5/3$, we can establish a linear relation between $\kappa$ and $q$ written as

\begin{eqnarray}
\label{rel}
\kappa=q-1.
\end{eqnarray}
In particular, the limits $\kappa=0, q=1$ and $\kappa=\frac{2}{3}, q=\frac{5}{3}$ lead to $G_{q}=G_{\kappa}$. It is clear that the relation $(\ref{rel})$ is valid only for the range $1\leq q < 5/3$.

Using Eq.(\ref{Gk}), in (\ref{lc}), we have derived the expression for the critical wavelength in the Kaniadakis statistics

\begin{eqnarray}
\label{lbdak}
\lambda_c^\kappa=\sqrt{\frac{(1+\frac{\kappa}{2})}{(1+\frac{3}{2}\kappa) \,2\kappa}\; \;
\frac{\Gamma{(\frac{1}{2\kappa}-\frac{3}{4})\,\Gamma{(\frac{1}{2\kappa}+\frac{1}{4})}}}{\Gamma{(\frac{1}{2\kappa}+\frac{3}{4})\,\Gamma{(\frac{1}{2\kappa}-\frac{1}{4})}}}}
\lambda_J.
\end{eqnarray}
This $\kappa$ modification of the Jeans' criterion leads to a new critical length $\lambda_c^\kappa$ that depends on the $\kappa$-parameter as follows:

\vskip .2cm
\noindent - If $\kappa=0 \Rightarrow \lambda>\lambda_c^\kappa=\lambda_J$, the usual Jeans criterion is recovered.
\vskip .1 cm
\noindent - If $0<\kappa<2/3$,  the Jeans criterion is modified as $\lambda>\lambda_c^\kappa>\lambda_J$.
\vskip .1 cm
\noindent - If $\kappa\rightarrow 2/3-, \lambda_c^\kappa\rightarrow \infty$, the self-gravitating system is always stable.
\vskip .2 cm

In Figure 1 we have plotted $\lambda_J^q$, Eq. (\ref{lbdaq}), and $\lambda_J^\kappa$, Eq. (\ref{lbdak}), both normalized by $\lambda_J$, Eq.(\ref{lc}), as function of $\kappa$. For this, we have used relation (\ref{rel}).

\begin{figure}
\begin{center}
\label{mass_q}
\includegraphics[width=3in, height=3in]{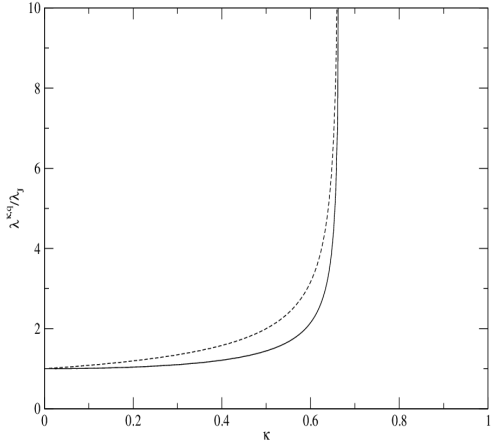}
\caption{Dashed line: Tsallis critical wavelength, $\lambda_J^q$. Solid line: Kaniadakis critical 
wavelength, $\lambda_J^\kappa$.}
\end{center}
\end{figure}

From Figure 1 we can observe that, except for the limit values, $\kappa=0$ and $\kappa=2/3$, the Tsallis critical wavelength is greater than the Kaniadakis critical wavelength. This result indicates that the NE effects of Tsallis' statistics lead to a self-gravitating system to be more stable when compared with the effects produced by Kaniadakis' statistics since $\lambda_J^q > \lambda_J^\kappa$.

It is possible to define a physical temperature \cite{DJ}, in the context of Tsallis' statistics. From Eqs.(\ref{ge}) and (\ref{lbdaq}) we can write

\begin{eqnarray}
\label{tne}
T_q=\frac{2T}{5-3q}.
\end{eqnarray}

It is easy to see that $T_q=T$ if $q\rightarrow 1$. Consequently we have the NE sound velocity

\begin{eqnarray}
\label{vne}
 v_{sq}=\sqrt{\frac{k_B T_q}{m}},
\end{eqnarray}
which agrees with the result Eq.(\ref{v2}).
We can also define a physical temperature in the context of Kaniadakis's statistics. From  Eqs.(\ref{keq}) 
and (\ref{lbdak}) we can write

\begin{eqnarray}
\label{tk}
T_\kappa=\frac{(1+\frac{\kappa}{2})}{(1+\frac{3}{2}\kappa) \,2\kappa}\; \;
\frac{\Gamma{(\frac{1}{2\kappa}-\frac{3}{4})\,\Gamma{(\frac{1}{2\kappa}+\frac{1}{4})}}}{\Gamma{(\frac{1}{2\kappa}+\frac{3}{4})\,\Gamma{(\frac{1}{2\kappa}-\frac{1}{4})}}}\;T.
\end{eqnarray}

We can show that $T_\kappa=T$  if $\kappa\rightarrow 0$. Consequently we have the $\kappa$-sound velocity

\begin{eqnarray}
\label{vsk}
 v_{s\kappa}=\sqrt{\frac{k_B T_\kappa}{m}},
\end{eqnarray}
which agrees with the result Eq.(\ref{vk2}).

\vskip 0.2 cm 

Let us now apply the results developed above and investigate how non-gaussian statistics can affect an astrophysical system.
We will use 16 mass measurements of high X-ray luminosity clusters in the redshift range of 0.17 to 0.55. 
The X-ray luminosity of galaxies provides up some information concerning their evolution, 
but here let us specifically analyze the masses of these 16 clusters obtained in \cite{data}. 
If the galaxies are in equilibrium, the virial mass of a cluster can be calculated as

\begin{equation}
 M = \frac{3}{G} \sigma_1^2 r,
\end{equation}
where $\sigma_1$ and $r$ are the velocity dispersion of the cluster and the three dimensional virial radius, respectively.
The virial mass can be rewritten as

\begin{equation}
\label{mass_q-eq}
 M_q = \frac{3}{G_q} \sigma_1^2 r\,\,,
\end{equation}
for the scenario modified via Tsallis statistics and

\begin{equation}
\label{mass_kappa-eq}
 M_{\kappa} = \frac{3}{G_{\kappa}} \sigma_1^2 r,
\end{equation}
for the framework modified by the Kaniadakis statistics. In Eqs. (\ref{mass_q-eq}) and (\ref{mass_kappa-eq}), $G_{q,\kappa}$ 
is given by Eq. (\ref{S}) and (\ref{Gk}), respectively.
\\

We will estimate from now on how the theoretical predictions given by Eqs. (\ref{mass_q-eq}) and (\ref{mass_kappa-eq}) can fit the data 
for the virial mass of the clusters compiled in \cite{data} from minimum chi-square estimation defined by

\begin{equation}
\label{mass_chi}
 \chi^2 = \sum_{i=1}^{16} \frac{(M_{obs} - M_{th})^2}{\sigma^2_M},
\end{equation}
where $M_{obs}$, $M_{th}$, and $\sigma_M$, represent the observation estimate for: the total mass  of the cluster, theoretical predictions
(Eqs. (\ref{mass_q-eq}) and (\ref{mass_kappa-eq})) and the error associated with observational measure, respectively.
\\

Figures 3 and 4 below
show the results of our statistical analysis as function of $\Delta \chi^2 = \chi^2 - \chi^2_{min}$ 
at 1$\sigma$ and 2$\sigma$ confidence level (CL) for the Tsallis and Kaniadakis statistics, respectively. 
We note that $q = 0.9918^{+0.0192}_{-0.0242}$ and $\kappa = 0.0119^{+0.0231}_{-0.0119}$ at 1$\sigma$ CL. During the accomplishment of the 
statistics, we considered $\kappa \geq 0$, in order to keep always $G_{\kappa} \geq 0$.
\\


Finally, we can say that in this work we have described both non-gaussian statistical formalisms, Tsallis and Kaniadakis, 
in the light of the star formation criterion formulated by Jeans at the beginning of the last century. 
We have seen that the limit for the Jeans instability condition is coherent with the standard values of $q$ and $\kappa$ in the literature.
After some computations, different values for the wavelength are compared trough comparing curves.  
We have investigated how non-gaussian effects can fit the masses of 16 galaxy clusters. 
We have found that the best fit value is completely compatible with BG statistics, 
but the non-gaussian effects cannot be discarded with the present set of data. More specifically we note,
(0.943) $0.968 \leq q \leq 1.01$ (1.035) and $0 \leq \kappa \leq 0.034$ (0.054) at 1$\sigma$ (2$\sigma$) CL. 

\begin{figure}
\begin{center}
\label{mass_q}
\includegraphics[width=3in, height=3in]{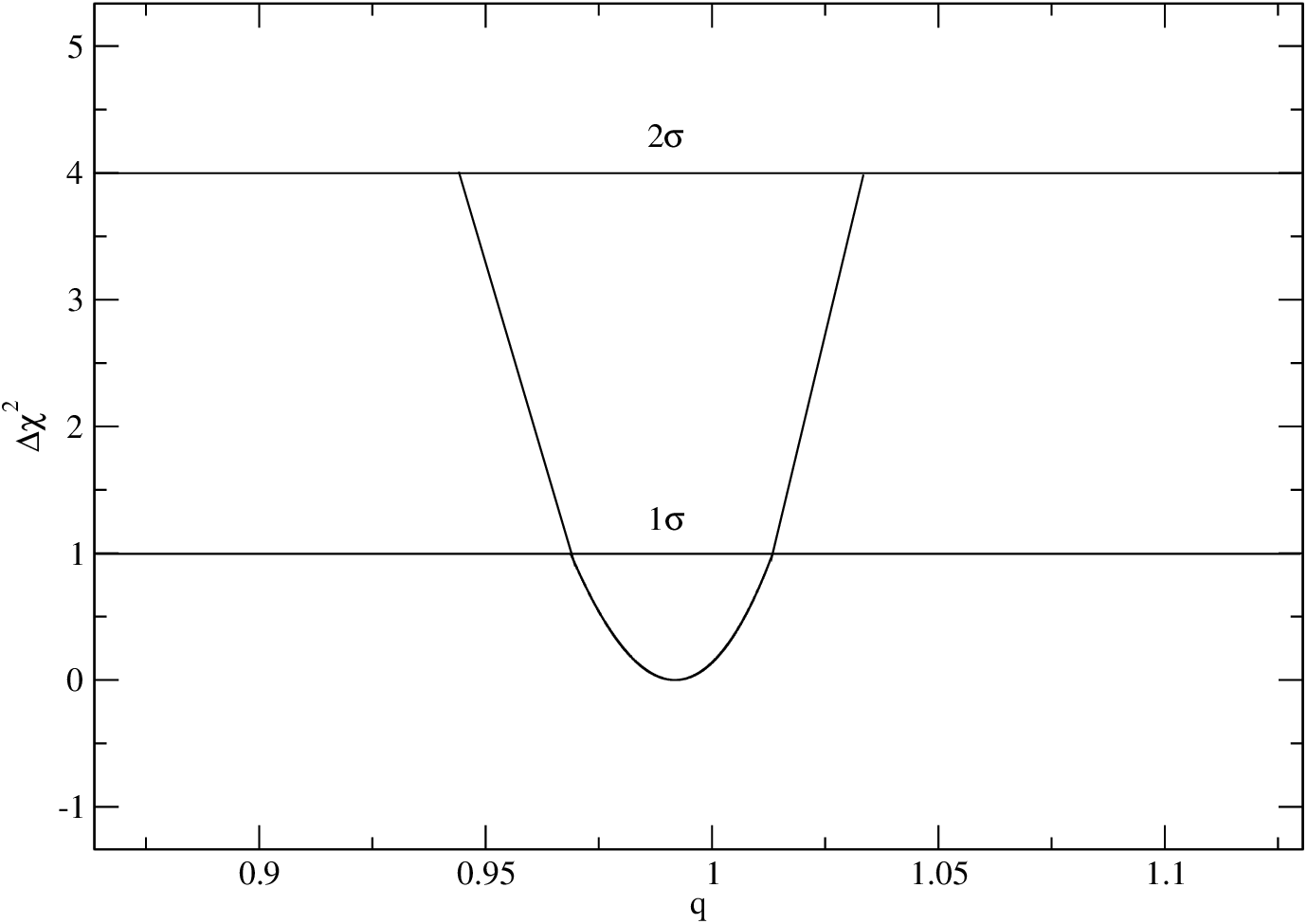}
\caption{The variance $\Delta \chi^2$ as a function of the parameter $q$.}
\end{center}
\end{figure}

\begin{figure}
\begin{center}
\label{mass_kappa}
\includegraphics[width=3in, height=3in]{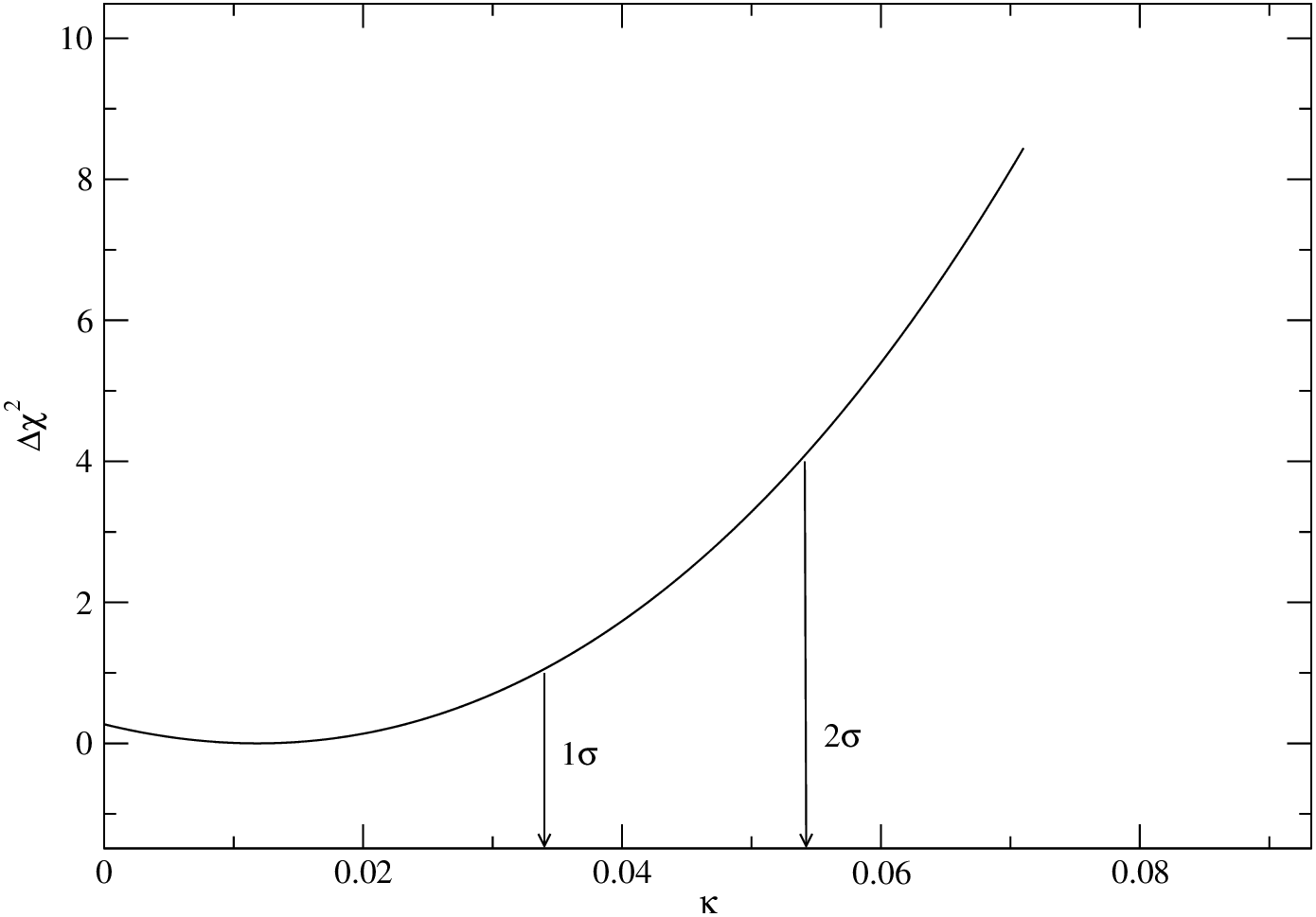}
\caption{The variance $\Delta \chi^2$ as a function of the parameter $\kappa$.}
\end{center}
\end{figure}

\acknowledgments

\noindent E.M.C.A. thanks the kindness and hospitality of Theoretical Physics Department at 
Federal University of Rio de Janeiro (UFRJ), where part of this work was carried out.
R.C.N. acknowledges financial support from CAPES (Coordena\c{c}\~ao de Aperfei\c{c}oamento de 
Pessoal de N\' ivel Superior), Scholarship Box 13222/13-9. 
CNPq and CAPES are Brazilian scientific support agencies.

\end{document}